\documentclass[twocolumn,%
 reprint,
 amsmath,amssymb,
 aps,
prd,
]{revtex4-1}
\pdfoutput=1

\usepackage[usestackEOL]{stackengine}
\usepackage{graphicx}
\usepackage{dcolumn}
\usepackage{bm}
\usepackage{bibentry}

\usepackage{enumerate}
\usepackage{nicefrac}
\usepackage{lipsum}

\usepackage{hyperref}
\usepackage{cleveref}

\newcommand{\bea}{\begin{eqnarray}}
\newcommand{\eea}{\end{eqnarray}}

\usepackage{adjustbox}

\usepackage{centernot}
\usepackage{color}
\newcommand{\be}{\begin{equation}}
\newcommand{\ee}{\end{equation}}

\makeatletter
\let\Hy@backout\@gobble
\makeatother

\begin{document}


\title{The Warped Dark Sector
}

\author{Philippe Brax,$\,^a$ Sylvain Fichet,$^{\,b,c}$ Philip Tanedo,$^{\,d}$}
\email{philippe.brax@ipht.fr, sfichet@caltech.edu, flip.tanedo@ucr.edu}

\affiliation{%
$^a$Institut de Physique Th\'{e}orique, Universit\'e Paris-Saclay, CEA, CNRS, F-91191 Gif/Yvette Cedex, France \\
$^b$Walter Burke Institute for Theoretical Physics, California Institute of Technology, Pasadena, CA
91125, California, USA \\
$^c$ICTP-SAIFR \& IFT-UNESP, R. Dr. Bento Teobaldo Ferraz 271, S\~ao Paulo, Brazil \\
$^d$ Department of Physics \& Astronomy, University of  California, Riverside, CA 92521
}

\begin{abstract}

Five-dimensional braneworld constructions in anti-de Sitter space naturally lead to dark sector scenarios in which parts of the dark sector vanish at high 4d momentum or temperature.
In the language of modified gravity, such feature implies a new mechanism for hiding light scalars, as well as the possibility of UV-completing chameleon-like effective theories.
In the language of dark matter phenomenology, the high-energy behaviour of the mediator sector  changes dark matter observational complementarity.
A multitude of signatures---including exotic ones---are present from laboratory  to cosmologic scales, including long-range forces with non-integer behaviour, periodic signals at colliders, ``soft bombs'' events well-known from conformal theories, as well as a dark phase transition and a typically small amount of dark radiation.

\end{abstract}

\maketitle

\section{Introduction}

Decades of astronomical observations point to the existence of dark matter (DM) and dark energy (DE).
It is a pressing question in fundamental physics to determine how these connect to the standard models of particle physics and cosmology.
Both DM and DE may suggest the existence of low-mass particles with weak interactions to visible matter.
The physics of such a \emph{dark sector} is the target of a new frontier of particle experiments~\cite{Battaglieri:2017aum,Alexander:2016aln}.

In this Letter we present a framework for a dark sector based on a truncated, warped extra dimension. Unlike the conventional Randall--Sundrum scenario, visible matter is localized on the ultraviolet (UV) brane. Dark particles are localized on the infrared (IR) brane and interact
 with the visible sector through a 5d bulk mediator particle. Recently it was shown that the IR brane becomes inaccessible to bulk fields with large absolute \emph{four}-momentum due to 5d gravitational dressing~\cite{mypaperI}.  As a result, high-energy experiments do not see the IR-localized particles and can only probe  the mediator's near-continuum.
A qualitatively similar behavior is  known to occur at finite temperature, where a phase transition replaces the IR brane by an AdS-Schwarzchild black hole~\cite{Creminelli:2001th}.

Through the AdS/CFT correspondence, our setting describes a strongly interacting and nearly-conformal sector coupled to the SM, which develops bound states at low-energy, \textit{i.e.} our  setup is the AdS dual of a \textit{composite dark sector} scenario.

Our dark sector scenario may be applied to constructing a theory of dark matter and  a theory of modified gravity.
 The scenario offers diverse and abundant observable phenomena; some are new, others have been mentioned in the vast extra dimensional and  CFT literature.  Our proposal sheds new light on these phenomena by connecting them in a unified theoretical framework and by making use of the precise quantitative developments which are possible in 5d.

\begin{figure}
\center
\includegraphics[width=7 cm,trim={0.cm 0cm 0.cm 0cm},clip]{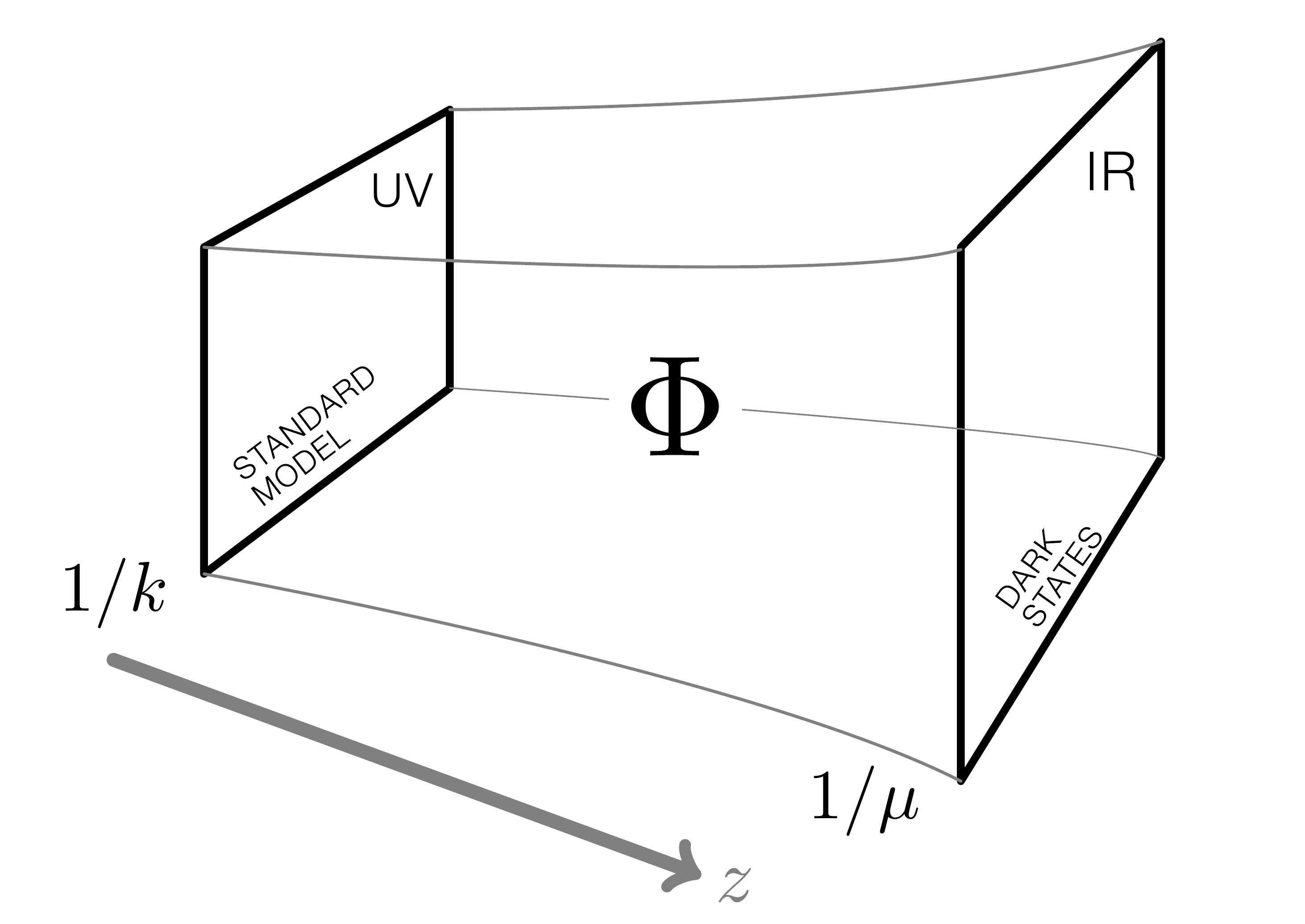}
\caption{  The warped dark sector:
 The ``visible'' sector (\textit{i.e.} the SM) lies on the UV brane of a slice of AdS$_5$, while
bulk and IR-brane localized degrees of freedom form the dark sector.
\label{fig:CDS}
}
\end{figure}

Dark sector constructions with a flat extra dimension have recently been proposed in~\cite{Rizzo:2018joy, Rizzo:2018ntg, Landim:2019epv}. Unlike these models, our scenario builds on properties that are unique to AdS space.
 Our proposal also differs from earlier warped constructions made in~\cite{vonHarling:2012sz,McDonald:2012nc,McDonald:2010fe,McDonald:2010iq}, and from the CFT portal of~\cite{Katz:2015zba}. Early work on 5d dark matter include~\cite{Cheng:2002ej,Servant:2002aq,Cembranos:2003mr,Agashe:2004bm,Agashe:2004ci, Rueter:2017nbk,Craig:2008vs,Craig:2008vs,Morrissey:2009ur}.


\section{AdS space and IR opacity}
\label{se:AdS}

The stabilization of curved, truncated background has been studied at length, see {e.g.}~\cite{Goldberger:1999uk}. We consider a slice of AdS space in the Poincar\'e patch with conformal coordinates, in which the metric is
\be
ds^2=\gamma_{MN}dX^MdX^N=(kz)^{-2}(\eta_{\mu\nu}x^\mu x^\nu-dz^2) \,
\ee
where $\eta_{\mu\nu}$ is Minkowski metric with $(+,-,-,-)$ signature.  The fifth dimension is assumed to be compact with $z\in[z_0,z_1]$, where $z_0\equiv 1/k$,   $z_1\equiv 1/\mu$ are respectively referred to as UV and IR branes. The AdS scale $k$ is taken to be $O(M_{\rm Pl})$ and the IR scale $\mu$ is a free parameter that may be very different than the TeV scale. We define the \emph{warp factor} $\varepsilon=\mu/k$.

We focus on a scalar field propagating in the 5d spacetime. The 5d action is
\be
\resizebox{\hsize}{!}{$\displaystyle
S=\int dX^M\left[\sqrt{\gamma} \left(\frac{1}{2}\nabla_M \Phi \nabla^M \Phi -\frac{1}{2} m_\Phi^2 \Phi^2 + {\cal L}_{\rm int}  \right) + \sqrt{\bar \gamma}{\cal L}_{\cal B} \right]
$}\,
\ee
where ${\cal L}_{\cal B}$ contains brane-localized Lagrangians and
 $\bar \gamma_{\mu\nu}$ is the induced metric on the branes, $\sqrt{\bar \gamma}=(kz)^{-4}$.
 We assume brane-localized mass terms,
\begin{align}
{\cal L}_{\cal B} =   -\frac{1}{2}\Phi^2\,  k  \big(
\delta_{z,z_0}\,(2-\alpha + b_{\rm UV})  - \delta_{z,z_1} \,(2-\alpha +  b_{\rm IR})
\big) \, \label{eq:5d_bound_action}
\end{align}
with $\delta_{u,v}\equiv \delta(u-v)$. The  special case $b_i=0$ for a given brane $i$ is compatible with the BPS condition $V_{\rm bulk}=(\partial V_i/\partial\phi)^2-V_i$, with $V_i$ the potential on brane $i$.  
The Feynman propagator is
\small\begin{align}
& \langle  \Phi(p,z)  \Phi(-p,z') \rangle   \equiv \Delta(p;z,z')= i\frac{\pi k^3 (zz')^2}{2 }  \times
\label{eq:propa_gen}
\\
 &
\frac{
\left[\tilde Y^{\rm UV}_{\alpha}J_{\alpha}\left(pz_<\right)
- \tilde J^{\rm UV}_{\alpha} Y_{\alpha}\left(pz_<
\right)\right]\left[
\tilde Y^{\rm IR}_{\alpha}J_{\alpha}\left(pz_>\right)
- \tilde J^{\rm IR}_{\alpha} Y_{\alpha}\left(pz_>
\right)
\right]}
{\tilde J^{\rm UV}_{\alpha} \tilde  Y^{\rm IR}_{\alpha}
- \tilde  Y^{\rm UV}_{\alpha} \tilde  J^{\rm IR}_{\alpha}}\, \nonumber
\end{align}
\normalsize
where $z_{<}={\rm min}(z,z')$, $z_>={\rm max}(z,z')$ and 
 $p=\sqrt{p^\mu p_\mu}$ is real for timelike  four-momentum $p^\mu$ and imaginary for a spacelike one. 
 The coefficients 
\be \tilde J^{\rm UV,IR}_{\alpha}=pk^{-1}J_{\alpha-1}(pk^{-1})-b_{\rm UV,IR}J_\alpha(pk^{-1})\ee
are  set by the boundary conditions. 
 Further  details on propagators  can  be found in {e.g.}~\cite{Ponton:2012bi, mypaperII}.

The propagator tends to be exponentially suppressed for $z_>\gg 1/|p|$ \textit{i.e.} in the IR region of the bulk, an AdS property with  no flat-space equivalent. 
 This behaviour has been long-known for spacelike momentum \cite{Goldberger:2002hb}. It has been recently shown that the suppression also occurs for timelike momentum, which is  relevant for $s$-channel processes   \cite{mypaperI}. This is a result of the dressing  from 5d interactions, including from 5d gravity,  \begin{align}
& \Delta^{\rm dr}(p;z,z')=\adjustbox{raise =0cm}{ \includegraphics[width=5. cm,trim={0.cm 0.cm 0.cm 0.3cm},clip]{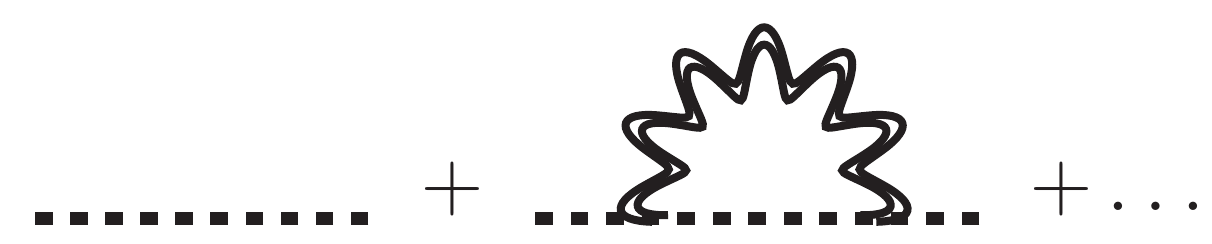}}
  \nonumber \\
& \approx \Delta(p(1+ i c);z,z') \propto  e^{- c pz_>}\,\quad {\rm if} \quad pz_>\gg 1\,.   \label{eq:propa_approx}
\end{align}

The $c$ coefficient, estimated in \cite{mypaperI}, comes from the imaginary part of the self energy resulting from the decay of the bulk scalar into AdS gravitational excitations  via the optical theorem.  
 For a first order phenomenological picture, it is enough to take $c$ as a  $ O(0.1-1)$ free parameter. 

 Interestingly, the exponential suppression plays the role of a censor for the IR region $p \gg 1/z_>$, which is the region where the 5d EFT breaks down \cite{Goldberger:2002cz, mypaperI}. 
From the viewpoint of the UV brane, an observer 
 producing $\Phi$ with timelike momentum sees a series of 
Kaluza-Klein (KK) modes
  becoming broader and broader and tending to a smooth continuum for $p\gg \mu$ (see Fig.~\ref{fig:PeriodicLHC}).

\section{The warped dark sector}
\label{se:model}

\begin{figure}
\center
\includegraphics[width=8.5 cm,trim={0.1cm 0cm 0.cm 0cm},clip]{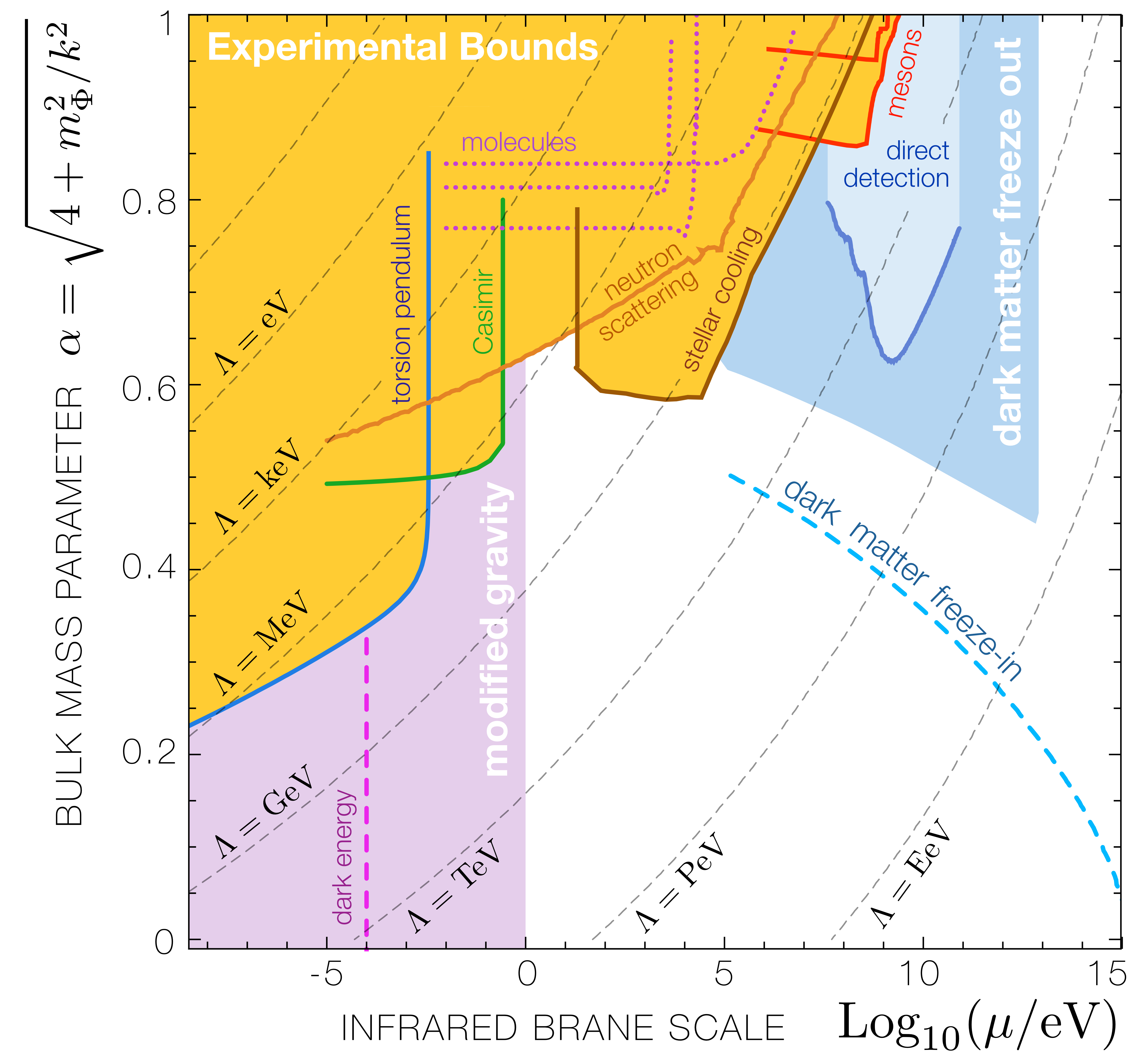} \\
\caption{
Scales and scenarios in the BPS (\textit{i.e.} $b=0$) case. Particles on the IR brane have $O(\mu)$ mass and $\Lambda$ is the SM-dark sector coupling, ${\cal L}\supset \frac{1}{\Lambda^2}{\cal O}_{\rm SM }{\cal O}_{\rm dark }$.  The orange region is excluded by measurements of processes involving the bulk mediator (see Sec.~\ref{se:signals}). Regions compatible with dark matter and modified gravity scenarios are also shown.
\label{fig:Scales}
}
\end{figure}

The fact that the IR region of the bulk becomes opaque for $p \gtrsim 1/z_>$ opens a tantalizing possibility of dark sector scenario.
 The SM may lie on the UV brane while additional dark particles, such as DM, may lie on the IR brane where the natural mass scale is $O(\mu)$.  
  The branes communicate via a bulk field  chosen to be a scalar $\Phi$ and described by the Green's function given in previous section. The action describing these interactions takes the form
\be \label{se:Sint_gen}
S\supset \int dX^M  \left(
\frac{\lambda}{\sqrt{k}}{\cal O}_{\rm SM}\Phi\delta_{z,z_0} +
\frac{\kappa}{\sqrt{k}}{\cal O}_{\rm D}\Phi\delta_{z,z_1}
\right)\,.
\ee
The IR-brane degrees of freedom and the bulk mediator together form the dark sector.

The key parameters in this framework are the dark sector scale $\mu$, the bulk mass parameter $\alpha$ which determines the KK mode profiles and thus the coupling to the SM brane, and the $b_{\rm UV}\equiv b$ parameter which is taken to be either $0$ (BPS case) or $O(1)$ (non-BPS case).  The other parameters are given values suggested by dimensional analysis, including $b_{\rm IR}=O(1)$.  For $b=O(1)$, the bulk field couples weakly to the UV brane. For $b=0$, this coupling can be larger.

Let us first study the low-energy behaviour, $p<\mu$. The bulk field is integrated out and the physics is described by a 4d EFT. When $b\neq 0$,   the effective operators have the following  magnitudes,
\be
{\cal L}_{4d}\sim \lambda^2 \frac{1}{k^2} \left( {\cal O}_{\rm SM}\right)^2
+
\lambda\kappa \frac{\varepsilon^{|\alpha|}}{k \mu} {\cal O}_{\rm SM}{\cal O}_{\rm D}
+
\kappa^2 \frac{1}{\mu^2} \left({\cal O}_{\rm D}\right)^2 \,. \label{eq:EFTbUV}
\ee
For $b=0$ and $\alpha<1$, the effective operators behave instead as
\be
{\cal L}_{4d}\sim \lambda^2 \frac{\varepsilon^{2-2\alpha}}{\mu^2} \left( {\cal O}_{\rm SM}\right)^2
+
\lambda\kappa \frac{\varepsilon^{1-\alpha}}{\mu^2} {\cal O}_{\rm SM}{\cal O}_{\rm D}
+
\kappa^2 \frac{1}{\mu^2} \left({\cal O}_{\rm D}\right)^2 \,. \label{eq:EFTbUV0}
\ee
For $b=0$ and $\alpha>1$, an ultralight mode  with mass $m_0\sim \mu \varepsilon^{\alpha-1}$ is also present in the spectrum, however this is not the focus of this letter. While the dark sector always has strong self-interactions, the effective SM-dark sector coupling ${\cal L}\supset \Lambda^{-2} {\cal O}_{\rm SM}{\cal O}_{\rm D} $ naturally ranges from strong ($\Lambda\sim \mu$) to  extremely small as a result of the localization of $\Phi$---which is controlled by $\alpha$.

At high energy, $p\gg\mu$, the IR region becomes opaque \cite{mypaperI}. The correlators between UV and IR branes are suppressed for both timelike and spacelike momentum,
\be
\langle
{\cal O}_{\rm SM} {\cal O}_{\rm D}
 \rangle \propto
\left\{
 e^{-c p /\mu}\,\, {\rm if}\,\, p^2>0\,,  \,\,\,
  e^{- p /\mu}\,\, {\rm if}\,\, p^2<0
\right\}\,.
\ee
This behaviour is totally different from a UV completion of the 4d EFT by  4d mediators, a standard  scenario in the dark sector literature. Here the number of 4d mediators is infinite and their couplings are  set by the theory such that destructive interference occurs, rendering the IR region inaccessible.

The  SM-to-SM amplitudes are also important for phenomenology. At $p\gg\mu$ they behave as if there was no IR brane,  for instance
\be
\langle
{\cal O}_{\rm SM} {\cal O}_{\rm SM}
 \rangle \propto
i\left[b k + (b+2\alpha)k\,\left(\frac{p}{2k}\right)^{2\alpha} \frac{\Gamma(-\alpha)}{\Gamma(\alpha)} \right]^{-1}
\, \label{eq:SMSM}
\ee
for $b\neq 0$, $\alpha>0$. This result can be exactly reproduced with a CFT model where the SM operator couples to a massive source $\phi_0$ mixing with a CFT operator ${\cal O}_{\rm CFT}$ of conformal dimension $\Delta=2+\alpha$%
\,\,  \cite{Gherghetta:2010cj,mypaperII}.

The \emph{warped dark sector}  can be envisioned as UV completion of low-scale 4d EFTs and is interesting as a DM or DE  scenario.  In this Letter we  focus only on a hadronic coupling, with ${\cal O}_{\rm SM}\equiv\bar N N$ below the QCD scale.
 As a direct consequence of opacity, the low-energy SM--dark sector coupling ${\cal L}_{4d}\supset \Lambda^{-2} {\cal O}_{\rm SM}{\cal O}_{\rm D} $  is only moderately restricted by experimental tests from the $p>\mu$ regime.
 The experimentally allowed values of $\Lambda$ are shown in Fig.~\ref{fig:Scales}. For example, the red giant bound vanishes below $\mu\sim 100 $\,eV as a result of opacity, such that a model with for instance $\mu\sim 10$\,eV, $\Lambda \sim 1 $\,MeV is \textit{not} excluded experimentally.  Improvements in molecular bounds or neutron scattering would be needed to probe such region.

\section{Modified gravity}
\label{se:modgrav}

The question of how to hide a light scalar has generated significant activity over the last decade \cite{Khoury:2010xi,Brax:2013ida}. A few mechanisms are known, such as the chameleon \cite{Khoury:2003aq} and Vainshtein mechanisms \cite{Babichev:2013usa}. Our setup introduces a new, geometric way to hide a scalar  $\varphi$: the IR brane where the scalar lives is ineluctably out of reach in the UV.
There are multiple model-building possibilities.

An attractive scenario is to have $f(R)$ gravity on the IR brane.  After   a Weyl transform this results in a scalar $\varphi$ coupled to the  stress-energy tensor of $\Phi$ evaluated on the IR brane. At low-energy the $\Phi$ fields can be integrated out at loop-level, leaving a 4d EFT containing $\varphi$ and the SM. 
  This interesting case is not treated further here.


A more minimal possibility is that the bulk field shares a bilinear term with a brane scalar ${\cal O}_{\rm dark}\equiv\varphi$, {i.e.}\ ${\cal L}\supset \delta_{z,z_1} k^{-1/2}\omega \Phi \varphi$ with $\omega=O(\mu^2)$. This case can be treated exactly by noting that $\varphi$ dresses the bulk field such that
\begin{align}
\label{eq:Deltadressed}
&\langle  \Phi(p,z)  \Phi(-p,z') \rangle= \Delta^{\rm dr}(p;z,z')- \\ & \Delta^{\rm dr}(p;z,z_1)\Delta^{\rm dr}(p;z_1,z')\frac{i\,\omega^2}{ p^2-m^2_{\varphi} + i\,\omega^2 \Delta^{\rm dr}(p;z_1,z_1)} \,. \nonumber
\end{align}
For $p<\mu$ the $\Phi$ propagators are constant and Eq.~\eqref{eq:Deltadressed} contains the light scalar pole with mass squared $\sim m^2_\varphi-i \omega^2 \Delta^{\rm dr}(m_\varphi;z_1,z_1)$ and a contact interaction---notice these features can be matched onto the 4d EFTs described in Eqs.~\eqref{eq:EFTbUV}-\eqref{eq:EFTbUV0}.
In contrast, at $p> \mu$ the $\Delta^{\rm dr}(p;z,z_1)$ propagators vanish exponentially,
such that only the $\Delta^{\rm dr}(p;z,z')$ term remains.
For instance, when considering the nonrelativistic spatial potential between ${\cal O}_{\rm SM}$ operators---given by $\langle  \Phi(p,z_0)  \Phi(-p,z_0) \rangle$,
Eq.~\eqref{eq:Deltadressed} interpolates between a $\sim 1/r$  potential at  distances $r>1/\mu$ and a short range, non-integer behaviour at $r<1/\mu$ (see details in Sec.~\ref{se:signals}).

Finally  our framework may also be used to UV complete models with a screening mechanism, such as chameleon or symmetron models (see \textit{e.g.} \cite{Khoury:2003aq,Hinterbichler:2010es,Burrage:2016bwy,Brax:2018grq}), that are otherwise ad-hoc low-energy EFTs with no obvious UV completion.

\section{Dark Matter}
\label{se:DM}

The observed abundance of DM can be explained by the existence of a sufficiently stable dark particle. Historically favored models of weakly interacting massive particles motivated by theories of electroweak naturalness are in tension with searches. A compelling alternative  is the secluded dark matter framework where DM interacts with visible matter through a low-mass mediator particle~\cite{Pospelov:2007mp, Alexander:2016aln}.  Such a possibility is naturally realized in our warped  framework, Eq.~\eqref{se:Sint_gen}, by identifying  ${\cal O}_{\rm D}=\bar \chi \chi$ where the DM particle $\chi$ is assumed to be a Dirac fermion localized on the IR brane.

A natural scale for the DM mass is $m_\chi \sim 4\pi \mu$ while the lightest  KK mode of the mediator, $\Phi$, has $O(\mu)$ mass. The $p$-wave $t$-channel annihilation of dark matter into the first KK mode(s) controls DM thermal production in the early universe and its subsequent present-day abundance. In turn, the properties of the mediator control the experimental signatures.
However, because of the properties of AdS space, these signatures differ significantly from those of standard 4d scalar mediators \cite{Krnjaic:2015mbs}. The resulting experimental fingerprint thus contrasts the benchmarks of experimental complementarity in dark matter searches.

There are essentially \textit{no} missing energy events at colliders for $|p|\gg\mu$.  The annihilation rate occurs in the intermediate energy regime $|p|\sim \mu$, while nucleon--DM scattering occurs in the 4d regime $|p|\ll \mu$ described by Eqs.~\eqref{eq:EFTbUV}-\eqref{eq:EFTbUV0}. The model's SM$\rightarrow \Phi \rightarrow$SM signatures are exotic, see Sec.~\ref{se:signals}.

Bulk field localization permits natural values of the mediator--SM coupling to over many orders of magnitudes, such that  both thermal freeze-out or freeze-in~\cite{PDG,Hall:2009bx,Krnjaic:2017tio} mechanisms may explain the abundance of DM thermal relics.
 Both regions are shown in Fig.~\ref{fig:Scales} in the BPS case. In the case of thermal freeze-out, the mass is bounded from above by annihilation unitarity~\cite{Griest:1989wd} and from below by dark radiation constraints~\cite{Boehm:2013jpa}. The coupling $\kappa$  is set to satisfy the DM abundance, which fixes direct detection bounds  shown in  Fig.~\ref{fig:Scales} \cite{PDG}. 
The first KK mode coupling to the SM depends crucially on its 5d profile, and cannot be too suppressed to maintain pre-freezeout thermal equilibrium \cite{Evans:2017kti}. 
  Freeze-in mostly depends on the mediator--SM coupling \cite{Blennow:2013jba,Krnjaic:2017tio} and thus on the first KK mode profile. The freeze-in mechanism can also take place in the non-BPS case for $\mu \gtrsim 10^3$~TeV.

Finally, the radion mode---whose exact mass depends on brane stabilization---can induce long range DM self-interactions while having a negligible coupling to the SM brane. Hence the warped dark sector naturally admits a mechanism for self-interacting DM as a way to address small scale structure~\cite{Tulin:2017ara}.

\section{Signatures and constraints}
\label{se:signals}

\begin{figure}
\center
\includegraphics[width=7 cm,trim={0.1cm 0cm 0.cm 0cm},clip]{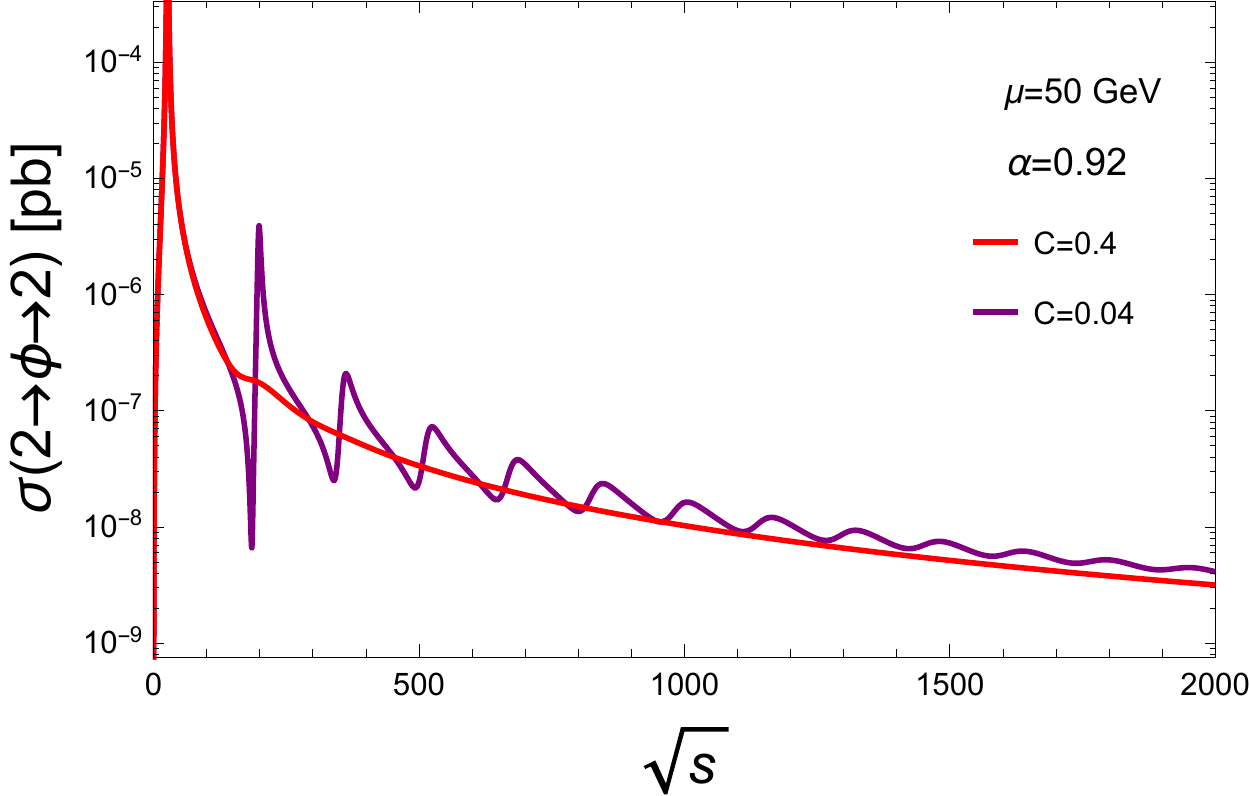} \\
\caption{
Example of periodic signal in SM $\rightarrow \Phi\rightarrow$ SM. Both bumps and dips are present. Interference with the SM is not shown and could dominate the signal.
\label{fig:PeriodicLHC}
}
\end{figure}

Our warped dark sector scenario has effects at terrestrial and astrophysical scales that qualitatively differ from scenarios with  4d mediators.

\underline{Non-integer fifth force}.
For $p>\mu$, the $t$-channel SM$\leftrightarrow$SM exchange of the KK modes induces  a non-relativistic long range potential between nucleons. Strikingly, this force has a non-integer power-law behaviour, such as
\be
V(r) \propto \frac{1}{r}\frac{1}{(kr)^{2-2\alpha}}
\ee
in the BPS case with $0<\alpha<1$. As a result, while fifth forces bound on the theory  are stringent for $\alpha\sim 1$, they are quickly relaxed for $\alpha<1$. Bounds on $V(r)$ obtained by recasting results from the E\"ot-Wash experiment \cite{Hoyle:2004cw}, Casimir measurements \cite{Chen:2014oda,Brax:2017xho}, molecular spectroscopy and neutron scattering data \cite{PhysRevD.87.112008,Salumbides:2013dua,Ubachs17, Nesvizhevsky:2007by,Fichet:2017bng} are shown in Fig.~\ref{fig:Scales}.

\underline{Momentum losses}.
Particles on the UV brane can  decay into states localized on the IR brane, such that UV observers see an effective loss of momentum. This usually puts strong bounds on  light dark sector models. 
 The situation in our warped dark sector is strikingly different because the energy loss into the dark brane is exponentially suppressed whenever $E_{\rm process}\gg \mu$.
The rates for such processes take an exact  form
\begin{align}
& \Gamma(\{SM\}\rightarrow \{SM'\} +(\Phi\rightarrow \{\rm dark\}))= \\ \label{eq:rate_inv}
&
\int \frac{dq^2}{\pi}
\Gamma_{\{SM\}\rightarrow \{SM'\} +\Phi}(q)
 \,|\Delta^{\rm dr}(q;z_0,z_1) |^2 \, q\,\Gamma_{\Phi\rightarrow \{dark\}}(q) \nonumber
\end{align}
where the suppression occurs from  $\Delta^{\rm dr}(q;z_0,z_1)$.
We show examples of exclusions from  red giants cooling  via Compton-like scattering \cite{Grifols:1986fc}  and from $K$ and $B$ mesons invisible  decays \cite{Artamonov:2008qb, Lees:2013kla, Willey:1982mc,Ball:2004ye, Knapen:2017xzo}. Collider bounds would require dedicated data analyses.

\underline{Soft bombs}.
Spherical soft   events with higher multiplicity are expected from both CFT \cite{Strassler:2008bv} and AdS sides \cite{Csaki:2008dt}.  Our model provides a concrete 5d realization  of the phenomenon.
The total rate has been analytically estimated in AdS in \cite{mypaperI} and seems to be exponentially suppressed with $p$. Nevertheless soft bombs are an important signature which should certainly be searched for.

\underline{Periodic signals at colliders}.
The KK near-continuum can be produced at colliders when the c.o.m energy exceeds $~\mu$.
 Production of a continuum is typically challenging to detect since its lineshape is similar to the one of the background.
Importantly, if the first KK modes are narrow enough, a periodic lineshape with bumps and dips is present. Such signature has been pointed out only recently in the context of the linear dilaton model \cite{Giudice:2017fmj}, our model reinforces the motivation for searching for such signature---temptingly by taking the Fourier transform of the signal. As an example we show a $q \bar q \rightarrow \Phi \rightarrow q \bar q$ cross section $\sigma\propto s |\Delta^{\rm dr}(\sqrt{s};z_0,z_0)|^2/k^4$ in Fig.~\ref{fig:PeriodicLHC}. 
 The presence of dips can be understood as a result of interferences between KK modes, and also happens in the interference with the background.

\section{High-temperature Cosmology}
\label{se:cosmo}

The warped dark sector at temperature $T\gtrsim \mu$ has two key features compared to 4d models.

\underline{Dark phase transition}.
First, it is well-known  \cite{Creminelli:2001th} that  a first order Hawking-Page-like phase transition occurs at  $T \sim \mu$. For $T>\mu$, the IR brane vanishes and the metric is AdS-Schwarchild, with a black hole in the IR region. This implies that the IR brane-localized states ({e.g.}~DM) do not exist in the $T>\mu$ phase, leaving only bulk excitations.  A key signature is that gravitational waves are generated  by nucleation during the phase transition, and could be accesssible by future gravitational waves experiments depending on the value of $\mu$ \cite{Schwaller:2015tja}.

\underline{Dark radiation}.
The second key feature is that dark radiation from the bulk mediators remains potentially quite small, unlike the effect of a simple relativistic 4d mediator.  This  fact has been studied in detail for gravitons in the original braneworld models \cite{Hebecker:2001nv,Langlois:2002ke,Langlois:2003zb}, and is tied to a subtle compensation between energy density and transverse pressure on the brane \cite{Langlois:2003zb}. The case of a bulk scalar turns out to depend crucially  on brane couplings and on $\alpha$. A detailed study is left for a dedicated work. Here we simply state that  for $\alpha<1/2$ the dark radiation behaviour is found to be similar to the graviton case, such that the effect of dark radiation is small enough to evade BBN bounds, thereby  allowing the dark sector scale $\mu$ to be below the MeV scale.

\section{Outlook}

We have argued that the properties of AdS have novel implications in the context of the dark sector paradigm. 
In this letter we present a general framework with specific choices. A number of developments follow from the present study, with for instance further analyses  of the opacity property, soft-bombs events, stellar cooling, dark radiation, or details of the dark phase transition.
The warped dark sector scenario opens new possibilities  for modified gravity  model-building and phenomenology, as well as for the secluded dark matter scenario. It also strengthens the case for a number of observables and new experiments, motivating  advances in Casimir and torsion pendulum experiments, neutron scattering and precision molecular spectroscopy,  as well as new kinds of LHC searches for soft bombs and periodic signals. 
The warped dark sector is also relevant for future experiments dedicated to dark sector searches \cite{Battaglieri:2017aum}.

 \section*{Acknowledgements}

We thank Nathaniel Craig, Csaba Cs\'aki, Gabriel Lee, Hai-Bo Yu, Stefania Gori, Gordan Krnjaic, Mark Wise, Prashant Saraswat,  Eric Perlmutter and David Poland for discussions.
PT is supported by the DOE grant DE-SC/0008541. PT thanks the Kavli Institute for Theoretical Physics (Grant No. NSF PHY-1748958) for its hospitality during the completion of this work.
  SF is supported by the S\~ao Paulo Research Foundation (FAPESP) under grants \#2011/11973, \#2014/21477-2 and \#2018/11721-4.   SF thanks Orsay University for hospitality and funding.
 This work is supported in part by the EU Horizon 2020 research and innovation programme under the Marie-Sklodowska grant No. 690575. This article is based upon work related to the COST Action CA15117 (CANTATA) supported by COST (European Cooperation in Science and Technology).


\section*{References}

\bibliography{biblio}

\end{document}


\vspace*{5mm}

\begin{center}
\huge Supplemental Material
\vspace{1cm}

\Large
Philippe Brax,$\,^a$ Sylvain Fichet,$^{\,b,c}$ Philip Tanedo,$^{\,d}$
\end{center}

\vspace{3cm}

\section{Limits of the propagator}

We present limits of the closed-form of the bulk field propagator. It is useful to quote the results for the rescaled field $\hat \Phi\equiv (kz)^{-1} \Phi $, in which case all metric factors on the branes are absorbed when brane fields are canonically normalized. The  propagator of the $\hat \Phi$ field is
\be
\hat\Delta(p;z,z') \equiv \langle  \hat\Phi(p,z)  \hat\Phi(-p,z') \rangle  = (kz)^{-1} (kz')^{-1} \Delta(p;z,z') \,.
\ee

\subsection{Low energy limit $|p|<\mu$}

In the $|p|<\mu$ \textit{i.e.}  4d limit,  we obtain
\be
k^{-1}\hat\Delta_p\left(\frac{1}{k},\frac{1}{k}\right)=-i
\frac{(2\alpha+b_{\rm IR})\left(\frac{k}{\mu}\right)^{2\alpha}-b_{\rm IR}}{
b_{\rm UV} (2\alpha+b_{\rm IR}) k^2 \left(\frac{k}{\mu}\right)^{2\alpha}
+b_{\rm IR} (2\alpha -b_{\rm UV}) k^2  
} \,,
\ee 
\be
k^{-1}\hat\Delta_p\left(\frac{1}{k},\frac{1}{\mu}\right)=-i
\frac{2\alpha\left(\frac{k}{\mu}\right)^{\alpha}}{
b_{\rm UV} (2\alpha+b_{\rm IR}) \mu k \left(\frac{k}{\mu}\right)^{2\alpha}
+b_{\rm IR} (2\alpha -b_{\rm UV}) \mu k 
} \,,
\ee
\be
k^{-1}\hat\Delta_p\left(\frac{1}{\mu},\frac{1}{\mu}\right)=-i
\frac{b_{\rm UV}\left(\frac{k}{\mu}\right)^{2\alpha}
+  (2+2\epsilon-b_{\rm UV})
}{
b_{\rm UV} (2\alpha+b_{\rm IR})  \mu^2 \left(\frac{k}{\mu}\right)^{2\alpha}
+b_{\rm IR}  (2\alpha -b_{\rm UV}) \mu^2
} \,.
\ee
The 4d effective operators described qualitatively   in Eqs.~(9),\,(10) of the Letter are  obtained exactly from the above expressions.

\subsection{High energy limit $k>|p|>\mu$}

 \subsubsection{UV-to-UV}

 We introduce
\be
S_\alpha=\frac{\sin\left(\frac{p}{\mu}-\frac{\pi}{4}(1-2\alpha)\right)}{\sin\left(\frac{p}{\mu}-\frac{\pi}{4}(1+2\alpha)\right)}
\ee

For $b_{\rm UV}\neq 0$ and  $\alpha>0$, we have 
\be
k^{-1}\hat\Delta_p\left(\frac{1}{k},\frac{1}{k}\right)=
i\left[b_{\rm UV} k^2 + (b_{\rm UV}+2\alpha)k^2\,\left(\frac{p}{2k}\right)^{2\alpha} \frac{\Gamma(-\alpha)}{\Gamma(\alpha)}S_\alpha \right]^{-1} \,.
\ee
   When $p/\mu$ has an imaginary part larger than $1$, we have  $S_\alpha\approx  (-1)^{\alpha}$.
For small  $b_{\rm UV}$, $b_{\rm UV}<|p|^2/k^2$, the cases $\alpha>1$, $0<\alpha<1$ have to be distinguished.    
For small $b_{\rm UV}$, for $\alpha>1$ we find 
   \be
k^{-1}\hat\Delta_p\left(\frac{1}{k},\frac{1}{k}\right)= i
\left[b_{\rm UV}k^2 +\frac{p^2}{2 (\alpha-1)}  + (b_{\rm UV}+2\alpha)k^2\,\left(\frac{p}{2k}\right)^{2\alpha} \frac{\Gamma(-\alpha)}{\Gamma(\alpha)}S_\alpha \right]^{-1} \,. \label{eq:Delta_UVUV_BPS}
\ee
For $0<\alpha<1$ and arbitrary $b_{\rm UV}$ we find
   \be
k^{-1}\hat\Delta_p\left(\frac{1}{k},\frac{1}{k}\right)=i
\frac{1}{(b_{\rm UV} +2\alpha)k^2} \frac{\Gamma(\alpha)}{\Gamma(-\alpha)} 
\left(\frac{2k}{p}\right)^{2\alpha} S_\alpha\,.
\ee

\subsubsection{UV-to-IR}

For  $b_{\rm UV}=O(1)$ and  $\alpha>0$, we have 
\be
k^{-1}\hat\Delta_p\left(\frac{1}{k},\frac{1}{\mu}\right)=
i\frac{\sqrt{\pi}}{b_{\rm UV}\Gamma(\alpha)}
\frac{1}{\sqrt{\mu^3 k}}\left(\frac{p}{2k}\right)^{\alpha-1/2} \frac{1}{\sin\left(\frac{p}{\mu}-\frac{\pi}{4}(1+2\alpha)\right)}\,.
\ee 
When $b_{\rm UV}$ is small ($|b_{\rm UV}|\lesssim|p|^2/(\alpha-1) k^2$), for $\alpha>1$ we have
\be
k^{-1}\hat\Delta_p\left(\frac{1}{k},\frac{1}{\mu}\right)= i
\frac{\sqrt{\pi}}{(b_{\rm UV} (\alpha-1)-\frac{p^2}{2k^2})\Gamma(\alpha-1)}
\frac{1}{\sqrt{\mu^3 k}}\left(\frac{p}{2k}\right)^{\alpha-1/2} \frac{1}{\sin\left(\frac{p}{\mu}-\frac{\pi}{4}(1+2\alpha)\right)}\,,
\ee 
while for $\alpha<1$ we have 
\be
k^{-1}\hat\Delta_p\left(\frac{1}{k},\frac{1}{\mu}\right)=- i
\frac{\sqrt{\pi}}{(b_{\rm UV}+2\alpha)\Gamma(-\alpha)}
\frac{1}{\sqrt{\mu^3 k}}\left(\frac{2k}{p}\right)^{\alpha+1/2} \frac{1}{\sin\left(\frac{p}{\mu}-\frac{\pi}{4}(1-2\alpha)\right)}\,.
\ee

  \subsection{Ultralight mode}

For $\alpha>1$ and $b_{\rm UV}$ small or zero, the spectrum contains an exponentially light mode of mass
 \be
m_0^2 = 2\beta b_{\rm UV} k^2+
\frac{2 b_{\rm IR} \mu^2 \beta (-b_{\rm UV}+2\beta+2) }{b_{\rm IR}+2+2\beta} \left(\frac{\mu}{k}\right)^{2\beta}
\label{eq:lightmode_bUV}
\ee   
where one has introduced $\beta=\alpha-1$. 
The mode remains light for \be
b_{\rm UV} \ll  \frac{\mu^2}{2\beta\,k^2} \,.
\ee

\section{Coupling to quarks and gluons}

Many observables considered in this Letter have energy scales below the QCD scale and can thus be described with an effective coupling to nucleons $\lambda_n$. 
 However for LHC collisions and meson decays the coupling to quarks and/or gluons needs to be specified.
Since we have a 5d effective theory, effective couplings to both quarks and to gluons should in principle be considered,
\begin{align}
& S\supset \int d^5x\, \delta_{z,z_0}\frac{\Phi}{\sqrt{k} \,M} \times \\
& \bigg(  -\lambda_q\sum_f   y^f_u \tilde H \bar u^f_R q^f_L-\lambda_q\sum_f  y^f_d  H \bar d^f_R q^f_L  + \lambda_G\alpha_s\left(G_{\mu\nu}^a\right)^2   \bigg) \,, \nonumber
\end{align}
 where $M$ is a typical scale in the SM sector, and can  for instance taken to be  $O(v)$. The couplings to  quarks are assumed to be  proportional to  quarks masses \textit{i.e.} minimally flavour-violating.
    In this Letter we do not consider LHC processes in details hence no further details are given. 
The meson decays are driven by the top-$W$ penguin diagram, which in our case is  proportional to $\lambda_t^2\equiv \lambda^2_q$.

The coupling to nucleons can be obtained from QCD considerations, by matching the QCD trace anomaly.  
 When, for instance, there are $n_h$ quarks which are heavy with respect to the absolute 4-momentum flowing in the effective nucleon-$\Phi$ vertex, the effective coupling reads
\be
\lambda_n=  \left(\frac{2}{27}\lambda_q - \frac{8 \pi \lambda_G}{9}\right) \frac{ m_N}{M} n_h \,.  \label{eq:Kratio} \,
\ee
This is obtained by integrating out the heavy quarks then matching the QCD trace anomaly.

For concreteness, we have taken $\lambda_n\sim 1$, $\lambda_t\sim 1$ in our analyses, which means we assume a large effective coupling to gluons $\lambda_G$.
 However we emphasize that, for smaller couplings (\textit{e.g.} a smaller $\lambda_G$), only the exclusion regions would somewhat shift, and all key results in the Letter remain unchanged.

\section{About 5d calculations}

\subsection{Kaluza-Klein representation}

In many of our calculations we use directly the closed-form representation of the propagator, Eq.~(4) of the Letter. However the Kaluza-Klein representation 
\be
\Delta_p(z,z')= i \sum_{n=0}^{\infty} \frac{f_n(z)f_n(z')}{p^2-m_n^2}\, \label{se:KK_dec}
\ee
 can also be useful,  because it provides a familiar 4d viewpoint on 5d physics.   The function  $f_n(z)$ is the profile of the $n$-th KK mode in the bulk. The $f_n(z)$ have dimension $1/2$. 

\subsection{Brane-localized dressing}

Brane-localized 1PI insertions dress the 5d propagator, and one can check that they end up modifying the boundary conditions. This can be seen  using either the dressed equation of motion, or directly using the geometric series representation of the dressed propagator. 
A consequence is that  the decay of the bulk field into brane-localized fields is simply encapsulated in an imaginary contribution to the brane mass terms.  That is, the brane self-energies can  be rigorously included in the propagator as a generalization of the $b_i$ parameters introduced in the Letter.

\subsection{Benchmark model }

The IR boundary condition in the plots shown in the Letter is taken to be $b_{\rm IR}=10+i\frac{p^2}{m^2_1}$ where $m_1$ is the first KK mode mass, which is roughly $m_1\sim 2 \mu$. The first KK mode is therefore a narrow resonance in this case.  
The bulk dressing is simply modeled by an imaginary part to $p$ starting near the second KK mode mass,  $\Delta(p(1+i\,c\, \Theta(p\sim m_2)\,),z,z')$~\cite{mypaperI}.

\section{Stellar cooling}

Here we review the stellar energy loss calculation leading to the bound shown in Fig.\,{\color{blue} 2} of the Letter.

\subsection{Basics}

The energy loss rate in the Boltzmann equation is given by
\be
L=\frac{1}{\rho} \int \frac{d^3 q_a}{(2\pi)^3} f_a(q_a) \int \frac{d^3 q_b}{(2\pi)^3} f_b(q_b) \, \sigma |v| \, E \label{eq:R_gen}
\ee
where $\rho$ is the medium density, $\sigma$ the cross-section of the process leading to energy loss, $|v|$ the relative initial velocity, and  $E$ is the energy carried away by the outgoing invisible state(s). Thermal equilibirum for the ingoing particles is assumed such that 
\be
f_{i}=g/(e^{(E_i-\gamma)/kT}\pm 1)
\ee
where $\gamma$ is the chemical potential and $g$ counts the degrees of freedom. When one of the particles, say $1$, has $m_1\gg T$, it is non relativistic and one has the simplification
\be
L\approx \frac{n_1}{\rho}\int \frac{d^3 q_b}{(2\pi)^3} f_b(q_b) \, \sigma |v| \, q_b 
\ee
Moreover, taking the other particle relativistic implies $|v|=1$. 

\subsection{Red giant data}

Following  \cite{Grifols:1986fc}, we consider the case of red giants, for which the anomalous energy loss rate is constrained to be 
\be
L_{\rm RG}^{\rm an} < 100~{\rm erg}\,\, {\rm s}^{-1} {\rm g}^{-1} \,.
\ee
A red giant is assumed to be mostly compound of Helium nuclei such that $n/\rho \sim 1/2$, and temperature is taken to be $T\sim 10^8$\,K.

\subsection{Energy loss rate into the warped dark sector}

As a typical example of stellar coupling bound we evaluate exactly the cooling of red giants from the Compton process
\be
\gamma N \rightarrow N + (\Phi\rightarrow \chi \bar\chi)
\ee
where the $\chi$ is a Dirac fermion localized on the dark brane and whose mass is neglected for simplicity. 
The process can be conveniently split into production and decay of $\Phi$ thanks to phase space recursion properties. This process can be directly computed in terms of 5d quantities, but it can be somewhat more intuitive to view it in terms of KK modes. The KK modes are denoted  $\phi^{(n)}$.    

The emission rate of an on-shell KK mode $\gamma N \rightarrow \phi^{(n)} N  $ from a Compton-like process in the non-relativistic limit is given by 
(using \cite{Grifols:1986fc})
\be
\sigma_{\rm Compton}^{(n)}(s,m_n^2)=  (f_n(z_0))^2 \frac{\kappa^2}{k} \frac{8\pi}{3} \frac{\alpha_e}{m_N^2}  \sqrt{1-\frac{m^2_n}{s}}\,.
\ee
The energy loss rate of the star from radiating the $\phi^{(n)}$ mode is then obtained from Eq.~\eqref{eq:R_gen}
\be
R^{(n)}(m^2_n)=\frac{1}{2m_N} \int^\infty_0
  \frac{dq q^2}{2\pi^2}\, \frac{1}{e^{q/T}-1} \, \sigma^{(n)}_{\rm Compton}(q^2,m_n^2)\, q\, .
\ee

Besides, the decay rate of $\phi^{(n)}$ into $ \chi \bar\chi$ is given by 
\be
\Gamma^{(n)}(m^2_n)= (f_n(z_1))^2 \frac{\lambda^2}{k}  \frac{m_n}{8\pi} \left(1-4\frac{m_\chi^2}{m^2_n}\right)\,. \label{eq:decay5dIR}
\ee 
Using phase space recursion one then obtains the energy loss of the (UV-localized) star into IR-brane localized states via the mode $\phi^{(n)}$,
\be
L^{(n)}(m^2_n) = 
\int_0^{\infty} \frac{dq^2}{\pi} 
R^{(n)}(q^2)
 \,\left|\frac{i}{q^2-m^2_n+i q \Gamma^{(n)}_{\rm tot}(q)}\right|^2  \, q\,\Gamma^{(n)}(q^2)  \,.
\ee
Finally the total energy loss rate into \textit{all} KK modes is  given by
\be
L=\sum_{n=0}^{\infty} L^{(n)}(m^2_n)\,. 
\ee
Here we make a slight abuse of notation for the sake of clarity, the KK modes self-energies are non-diagonal and should be treated as a matrix in the KK representation of the propagator \cite{mypaperI}. This approximation will be removed in the next steps. 
Let us introduce the 5d energy loss rate $R(q^2)$ and the 5d decay width  $\Gamma_{\Phi\rightarrow \chi \bar \chi}(q^2)$,  satisfying
\be
R^{(n)}(q^2) \equiv (f_n(z_0))^2  R(q^2)\,,
\ee
\be
\Gamma^{(n)}(q^2) \equiv (f_n(z_1))^2 \Gamma_{\Phi\rightarrow \chi \bar \chi}(q^2)\,.
\ee
It immediately follows from the KK representation Eq.~\eqref{se:KK_dec} that the total energy loss rate is given by
\be
L = 
\int_0^{\infty} \frac{dq^2}{\pi} 
R(q^2)
 \,|\Delta(q;z_0,z_1) |^2 \, q\,\Gamma_{\Phi\rightarrow \chi \bar \chi}(q^2) \nonumber \,. \label{eq:RGlossrate}
\ee
This result can be equivalently obtained from 5d Feynman rules.

Imposing $L\leq L_{\rm RG}^{\rm an}$ gives an exclusion region shown in Fig.\,{\color{blue} 2} of the Letter.

\section{Meson decays}

Collider measurements of heavy meson decays put stringent bounds on light dark sectors, the strongest being from $B$ and $K$ mesons. Experimental bounds are
\be
{\rm BR}(B\rightarrow K + {\rm inv})< 1.6 \cdot 10^{-5}\,, \label{eq:Bex}
\ee
\be
{\rm BR}(K\rightarrow \pi + {\rm inv})= 1.73^{+1.15}_{+1.05 } \cdot 10^{-10}\, \label{eq:Kex}
\ee
from \cite{Artamonov:2008qb,Lees:2013kla}. 

The main contribution to invisible decays comes from the penguin diagram with a top/$W$ loop. The meson decay rates are thus proportional to the $\Phi$ coupling to the top quark, ${\cal L}\supset \kappa_t\bar t t \Phi\,\delta(z-z_0)$. The intermediate KK continuum is in principle highly unstable and should end with narrow states on the dark brane. The processes to consider are thus
\be
B\rightarrow K + (\Phi\rightarrow \chi \bar \chi) \,, \quad 
K\rightarrow \pi + (\Phi\rightarrow \chi \bar \chi) \,.
\ee
As for stellar cooling, the exact decay rates can be found starting from existing 4d results. Using phase space recursion, the $B$ meson decay rate is found to be 
\be
\Gamma_{B\rightarrow K + \chi \bar \chi} = 
\int_0^{(m_B-m_K)^2} \frac{dq^2}{\pi} 
\Gamma_{B\rightarrow K + \Phi}(q^2)
 \,|\Delta(q;z_0,z_1) |^2 \, q\,\Gamma_{\Phi\rightarrow \bar \chi \chi}(q^2) \nonumber \,, \label{eq:RGlossrate}
\ee
where the intermediate decay rate into the KK continuum, deduced from  \cite{Knapen:2017xzo,Willey:1982mc}, is given by
\be
\Gamma_{B\rightarrow K + \Phi}(q^2)= \frac{\kappa_t^2}{k}
 \left(\frac{m^2_B-m_K^2}{m_b-m_s}\right)^2\left(\frac{3m_b m^2_t V_{ts}V_{tb}}{16\pi^2 v^3}\right)^2\, \frac{K(m_B^2;m_K^2,q^2)}{16 \pi m_B}F_0(q^2)
\,
\ee
with the kinematic factor
\be
K(M^2;m_1^2,m_2^2)=\frac{\sqrt{(M^2-(m_1+m_2)^2)(M^2-(m_1-m_2)^2)}}{M^2}\,
\ee
and the form factor \cite{Ball:2004ye}
\be
F_0(q^2)= 0.33 \left(1-\frac{q^2}{38{\rm GeV}^2}\right)\,.
\ee
The $\Gamma_{\Phi\rightarrow \chi \bar \chi}$ rate is given by Eq.~\eqref{eq:decay5dIR}.

The kaon decay is given by 
\be
\Gamma_{K\rightarrow \pi + \Phi}(q^2)= \frac{\kappa_t^2}{k}
 \left(\frac{m^2_K-m_\pi^2}{m_s-m_d}\right)^2\left(\frac{3m_s m^2_t V_{td}V_{ts}}{16\pi^2 v^3}\right)^2\, \frac{K(m_K^2;m_\pi^2,q^2)}{16 \pi m_K}\,
\,
\ee
and a calculation analog to the $B$ meson case can be done. 

We obtain the exclusion regions in Fig.\,{\color{blue} 2} of the Letter by comparing the anomalous  decay rates with the experimental bounds from Eqs.~\eqref{eq:Bex},\,\eqref{eq:Kex}.

\section{SM$\rightarrow $SM cross section}

The cross sections shown in Fig.\,{\color{blue} 3} of the Letter are given by 
\be
\sigma(s)=\frac{s}{4\pi}  \,\frac{\lambda^4}{k^2} \left|
\Delta^{\rm dr}(\sqrt{s};z_0,z_0)
\right|^2
\ee
and taking $\lambda=1$. At a collider, such distribution would be seen  by plotting the invariant mass of the final states. 
 Periodic bumps are always present but their exact shape and smearing depends on bulk and brane dressing. We  have used $b_{\rm IR}=10+i\frac{p^2}{m^2_1}$ and $c=0.4,0.04$ in the example shown in Fig.\,{\color{blue} 3}.

\section{Neutron scattering}

Following the analyses of \cite{Fichet:2017bng,  Brax:2017xho},  we constrain the neutron-neutron interaction using the optical method. The neutron scattering length is defined as 
\be
\sqrt{\frac{\sigma({\bf q})}{4\pi}}=l({\bf q})\,
\ee
where one has $l({\bf q})= 2m_N \tilde{V}({\bf q}) $, where  $\tilde{V}({\bf q})$ is the scattering potential. 
The experimental bound obtained by combining total cross section and optical measurements sets \cite{Nesvizhevsky:2007by}
\be
l(k_{\rm ex})-l(0) < 6\cdot 10^{-4}\,\,{\rm fm}\, \label{eq:neutron_ex}
\ee
with $k_{\rm ex}=40$\,keV.
In our model the scattering length is set by
\be
l({\bf q})= -i 2m_N \Delta_{i q}(z_0,z_0) \,.
\ee
Using the constraint Eq.~\eqref{eq:neutron_ex} on $l({\bf q})$ then 
 gives the exclusion region shown in Fig.\,{\color{blue} 2} of the Letter.

\section{Long-range force}

Let us start with the $t$-channel diagram $i{\cal M} = N  N \rightarrow  N N $ induced by  $\Phi$ exchange. 
 The (spacelike) exchanged momentum is denoted $p$. 
  The non-relativistic scattering potential is given by (we follow the conventions of \cite{Brax:2017xho})
\be
i{\cal M} = -i \tilde V({\bf |q|})   4m_N^2 \delta^{s_1s_1'}\delta^{s_2s_2'}\,, \label{eq:NRV}
\ee
where $p^2=-\bf |q|^2$.
In the BPS case we have therefore
\be
\tilde V({\bf |q|}) = -i \frac{\kappa \lambda}{k} \Delta_{\bf |q|} (z_0,z_0)
= -\frac{ \kappa \lambda}{2 k^2}\frac{\Gamma(\alpha)}{\Gamma(1-\alpha)}\left(\frac{4k^2}{{\bf |q|}^2}\right)^\alpha\,.
\ee
The spatial potential is given by the Fourier transform
\be
V(r)=\int\frac{ d^3 { \bf q} }{(2\pi)^3} \tilde V(|{\bf q}|) e^{i { \bf q}\cdot {\bf r}}\,
\ee
where $|{\bf r}|=r$. 

The integral can be evaluated using the Schwinger trick, 
\begin{align}
 \int\frac{ d^3 { \bf q} }{(2\pi)^3} &  {\bf |q|}^{-2\alpha} e^{i {\bf q}.r}    =
\int\frac{ d^3 { \bf q} }{(2\pi)^3} \frac{1}{\Gamma(\alpha)} \int dt\,t^{\alpha-1} {\bf |q|}^{-2\alpha} e^{-t{\bf q}^2} e^{i {\bf q}.r}\\ \nonumber
&=  \frac{1}{8\pi^{3/2}\,\Gamma(\alpha)} \int dt\,t^{\alpha-5/2} {\bf |q|}^{-2\alpha} e^{-\frac{r^2}{4t}} 
=  \frac{1}{8\pi^{3/2}\,\Gamma(\alpha)} \int d\xi\,\xi^{1/2-\alpha} {\bf |q|}^{-2\alpha} e^{- \xi
\frac{r^2}{4}}\\  \nonumber
& =  \frac{\Gamma(3/2-\alpha)}{8\pi^{3/2}\,\Gamma(\alpha)} \left(\frac{2}{r}\right)^{3-2\alpha}\,,
\end{align}
and the spatial potential reads
\be
V(r) = - \frac{\kappa \lambda }{2 \pi^{3/2}}\frac{\Gamma(3/2-\alpha)}{\Gamma(1-\alpha)} \frac{1}{r} 
\left(\frac{1}{kr}\right)^{2-2\alpha} \,.
\ee

This potential can then be used in  molecular, Casimir and torsion pendulum measurements following Refs.~\cite{Fichet:2017bng, Brax:2017xho}, which gives exclusion regions shown in Fig.\,{\color{blue} 2}  of the Letter.

\section{Dark radiation}

 The Boltzmann equation for the energy density on the UV brane reads \cite{Langlois:2002ke}
  \be
  \frac{d\rho}{dt}+ 3 H (\rho + p) = -\sigma \equiv -\sum_n \sigma_n= -\sum_n
  \int \frac{d^3 p_n}{(2\pi)^3} {\bf C}_n \, \label{eq:Boltz_KK}
  \ee 
   with
   \be
   {\bf C}_n= \frac{1}{2} \int \frac{d^3 p_1}{(2\pi)^3\, 2E_1}
   \int \frac{d^3 p_2}{(2\pi)^3\,2 E_2} \overline{|{\cal M}_n(s)|^2} f_1 f_2 (2\pi)^4 \delta^{(4)}(p_1+p_2-p_n)\,.
   \ee
   The sum in Eq.~\eqref{eq:Boltz_KK} is over KK modes.
Performing the $d^3p_n$ integral and the angular integrals which remove the remaining delta, on obtains
\be
\sigma_n= \frac{1}{32 \pi^3}
 \int d E_1 \int d E_2 (E_1+E_2) \,  \overline{|{\cal M}_n(m^2_n)|^2}\,f_1 f_2 \, \Theta(m_n^2<4 E_1 E_2)\,.
\ee 
   We then have
   \be
   \sigma=\frac{1}{32 \pi^3}
 \int d E_1 \int d E_2 (E_1+E_2)\,f_1 f_2 \, \sum_{n=0}^{\tilde n} \overline{|{\cal M}_n(m^2_n)|^2} \, 
   \ee
where $\tilde n$ denotes the threshold,  $m^2_{\tilde n}\approx 4E_1 E_2$.

In the scalar case, SM fermions annihilations into a given scalar KK mode  $\Psi \bar \Psi \rightarrow \Phi^{(n)}$ is given by the rate
\be
\sigma_n=\frac{\lambda^2}{k}     f^2_n(z_0)\, 2 m_n^2\,
\ee
for a Yukawa coupling $\lambda$.
Using the contour trick of \cite{mypaperI}, we have
\begin{align}
&\sum_{n=0}^{\tilde n} f^2_n(z_0) m_n^2    =\frac{1}{-2\pi} \int_{{\cal C}[\tilde n]}d\rho\,
\Delta_{\sqrt{\rho}}(z_0,z_0) \rho \\ \nonumber & =
(4E_1 E_2)^2 \left(\frac{k^2}{E_1 E_2}\right)^\alpha\frac{1}{(b_{\rm UV}+2 \alpha)k} \frac{1}{\Gamma(1-\alpha)\Gamma(-\alpha)(\alpha-2)} \equiv  \frac{\tilde C}{k} (E_1 E_2)^2 \left(\frac{k^2}{E_1 E_2}\right)^\alpha \,
\end{align}
where we have considered the BPS case with $0<\alpha<1$ and used the corresponding limit Eq.~\eqref{eq:Delta_UVUV_BPS} for the propagator.  

Performing the thermal integrals, the rate loss into the bulk scalar is found to be
\be
\sigma=  \frac{\tilde C}{8\pi^3}   \Gamma(3-\alpha)\Gamma(4-\alpha) L_{3-\alpha}(1)L_{4-\alpha}(1)\, \lambda^2\,T^5 \left(\frac{T }{k}\right)^{2-2\alpha}\,,
\ee
and thus   
\be
\sigma=  \frac{2\alpha(3-\alpha)(2-\alpha)(1-\alpha)^2  L_{3-\alpha}(1)L_{4-\alpha}(1)}{\pi^3(b_{\rm UV}+2\alpha)}   \, \lambda^2\,T^5 \left(\frac{T }{k}\right)^{2-2\alpha}\,.
\ee   

A simple estimate \`a la \cite{Hebecker:2001nv} is possible only for $\alpha<1/2$, in which case the 
dark radiation fraction
\be
\Omega_D \approx \int_{\tau_i}^\infty d\tau \,\frac{\sigma}{\rho_{\rm tot}} 
\ee
with $\rho_{\rm tot}= \frac{\pi^2}{30} g_*  T^4  $, $T^2=\frac{90}{2\pi}\frac{g_*M_{\rm Pl} }{\tau}$,
is dominated by early times as in the familiar graviton case of usual braneworld models.

\section{Dark Matter}

\subsection{Freeze-out scenario}

The  annihilation cross section of Dirac dark matter into the first KK mode  $\chi \bar \chi \rightarrow \Phi^{(1)}\Phi^{(1)}$ is given by 
\be
\langle \sigma v \rangle =\frac{\left(\kappa f_1(z_1)\right)^4}{k^2} \frac{3 v^2}{128\pi m_\chi^2}\,,
\ee
with $v^2\sim 3T/2m_\chi\sim 0.3$\,. Moreover the DM abundance from thermal freeze-out is predicted to be \cite{PDG}
\be
\Omega h \sim \frac{0.2\,{\rm pb}}{\langle \sigma v \rangle}\,.
\ee
The observed value is
\be
\Omega h_{\rm obs} \approx 0.12\,,
\ee
thereby giving the constraint
\be
\left(\frac{ \kappa f_1(z_1) }{0.4 \sqrt{k}}\right)^4
\left(\frac{100 {\rm GeV}}{m_\chi}\right)^4\approx 1\,.
\ee
This equation sets $\kappa$ as a function of the warped dark sector parameters $\mu$, $\alpha$. 

The value of $\kappa$ obtained is then used to predict the DM-nucleon scattering cross section needed for direct detection. The freeze-out region and direct detection bound  are shown in Fig.\,{\color{blue} 2} of the Letter.

\subsection{Freeze-in scenario}

Freeze-in mechanism assumes vanishing dark matter abundance at early times \cite{Hall:2009bx}. Dark matter is slowly produced via a small but non vanishing coupling to the thermal bath, which in our case naturally results from the exponentially suppressed first KK mode profile on the UV brane, $f_1(z_0)$. 
In the case of a mediator lighter than the DM particle, following the analysis of \cite{Blennow:2013jba,Krnjaic:2017tio}, DM abundance is roughly independent on the DM mass and rather depends on the mediator coupling to the SM. In our model we obtain
\be
\lambda f_1(z_0) \sim 10^{-11} \,. 
\ee
This gives a line in the $\mu-\alpha$ plane which is shown in Fig.\,{\color{blue} 2} of the Letter. 

\bibliographystyle{IEEEtran}
%
